\documentclass[11pt]{article}
\def\lb{\label}
\def\be{\begin{equation}}
\def\ee{\end{equation}}
\def\ba{\begin{eqnarray}}
\def\ea{\end{eqnarray}}
\def\part{\partial}

\def\tr{\tilde{r}}

\begin{document}

\begin{center}
{\bf {\Large Analytical approach to critical scalar field
collapse\\ in three dimensions}} \\
\bigskip
{\bf G\'erard Cl\'ement}$^{a}$
and {\bf Alessandro Fabbri}$^{b}$ \\
\medskip
{\small $^{(a)}${\it Laboratoire de  Physique Th\'eorique LAPTH (CNRS),}} \\
{\small {\it B.P.110, F-74941 Annecy-le-Vieux cedex, France}}\\
{\small $^{(b)}${\it Dipartimento di Fisica dell'Universit\`a di
Bologna and INFN sezione di Bologna,}} \\
{\small {\it Via Irnerio 46,  40126 Bologna, Italy}}
\end{center}

\bigskip
\begin{abstract}
In the quest of the critical solution for scalar field collapse in
2+1 gravity with a negative cosmological constant, we present a
one parameter family of solutions with continuous self similar
(CSS) behaviour near the central singularity. We also discuss
linear perturbations on this background, leading to black hole
formation, and determine the critical exponent.
\end{abstract}
\bigskip

Recent  numerical investigations \cite{CP,HO} of near-critical scalar field collapse in 2+1 dimensional AdS
spacetime show a continuous self-similar (CSS) behaviour near the central singularity and power law scaling for
the black hole mass. Garfinkle \cite{gar} found that this behaviour is well approximated, at intermediate times,
by a member of a one-parameter family of exact CSS solutions to the equations with vanishing cosmological constant
$\Lambda$. However the extension of these solutions to $\Lambda\neq 0$, necessary to explain black hole formation
(as vacuum black holes exist only for $\Lambda < 0$ \cite{BTZ}) is a hard problem which has not been solved up to
now.

By a limiting process, we have derived from these solutions a new class of CSS solutions which can be extended to
$\Lambda\neq 0$. These solutions
\ba\label{new2} ds^2 & = & dudv - (-u)^{2/(1+c^2)} d\theta^2,
\nonumber \\ \phi & = & -\frac{c}{1+c^2}\ln(-u) \ea present for $c^2 \le 1$ a point singularity ($u=0$,
$v=+\infty$), and for $c^2 < 1$ a null line singularity $u = 0$. They can be extended to exact solutions of the
$\Lambda = -l^{-2} < 0$ equations by means of the ansatz
\be\lb{anext} ds^2 = e^{2\sigma(x)}dudv - (-u)^{\frac2{c^2+1}}\rho^2(x)d\theta^2, \quad \phi =
-\frac{c}{c^2+1}\ln|u| + \psi(x)\,,
\ee where $x = uv$, with the initial conditions  $\rho(0) = 1, \, \sigma(0) = \psi(0) = 0$. These extended
solutions inherit the CSS behaviour close to the central singularity, and have the correct AdS behaviour at
spatial infinity.

To show that these quasi-CSS solutions are indeed threshold solutions for black hole formation, we study their
linear perturbations. Expanding these in modes $(-u)^{-k/(1+c^2)}$ and keeping only one mode, the perturbed
perimeter function $r = (-g_{\theta\theta})^{1/2}$ reads (similar expressions hold for $\sigma$ and $\phi$)
\be\lb{pertan} r = (-u)^{1/(c^2+1)}(\rho(x) + (-u)^{-k/(c^2+1)}\tr(x)). \ee
The eigenmodes are determined by enforcing appropriate boundary conditions: (A) on the line $u = 0$, the perturbed
metric should not diverge too quickly; (B) on the line $v=0$ ($x = 0$) the perturbed solution should smoothly
match the background ($\tilde r(0)=\tilde \sigma (0)=\tilde \phi (0)=0$) and be extendible (i.e. analytical); (C)
the perturbations should not blow up on the AdS boundary. It turns out that this last condition is always
satisfied up to gauge transformations.

The case $c=0$ is at the same time simple and interesting. In this case the scalar field decouples, $\phi=0$, and
the corresponding quasi-CSS solution is the vacuum BTZ \cite{BTZ} solution. The linear perturbation equations with
our boundary conditions lead to a unique solution, with to the eigenvalue $k=2$. This solution is found to be an
exact linearization in $M$ of the BTZ black hole, written as
\be ds^2 = \rho(dudv-u^2d\theta^2), \quad \rho = (1+uv/4l^2)^{-2},
\ee with the perimeter function $ r= -u(\rho +\frac{M}{4}u^{-2}x\rho)$.

In the case of genuine scalar perturbations, our boundary conditions lead to two possible modes $k_a = c^2 + 3/2$
and $k_b = c^2 + 2$. For the $a$ mode the singularity and the apparent horizon appear simultaneously on the null
line $u = 0$ at the time $v=0$ and evolve in the region $v>0$ in a physically meaningful way, while for the $b$
mode the singularity still appears for $v = 0$, but the apparent horizon seems to be eternal, as in the case of
the static BTZ black hole. We argue that this last solution, which cannot describe actual gravitational collapse
with regular initial conditions, should be excluded as unphysical, leaving only one growing mode $a$.

The critical exponent is related to the eigenvalue $k$ by $\gamma=1/k$. In the BTZ case $c=0$ we obtain
$\gamma=1/2$. In the scalar field case, choosing the critical value $c^2 = 1$ we obtain for the $a$ mode $\gamma =
0.4$. This value differs signicantly from those found in the numerical analysis, i.e. $\gamma \sim 1.2$ in
\cite{CP} and $\gamma\sim 0.81$ in \cite{HO}, showing that further work is therefore needed in order to clarify
this issue.

\end{document}